# FINITE ELEMENT ANALYSIS OF THE INDENTATION TEST ON ROCKS WITH MICROSTRUCTURE

Jean Sulem* ,CERMES - Ecole Nationale des Ponts et Chaussées / LCPC, France

Miguel Cerrolaza, Facultad de Ingenieria, Universidad Central de Venezuela

**Abstract**

Strength parameters of rocks are currently determined from indentation tests. In this paper, a finite element analysis of this test is presented and scale effect is studied. The rock is modelled as an elasto-plastic medium with Cosserat microstructure and consequently possesses an internal length. The response of the indentation curve is studied for various values of the size of the indentor as compared to the internal length of the rock in order to assess the scale effect.

---

* **Corresponding author**.

Postal address : CERMES, Ecole Nationale des Ponts et Chaussées, 6 et 8 avenue Blaise Pascal, Cité Descartes, Champs-sur-Marne, 77455, Marne-La-Vallée, Cédex 2, France.

Phone : +33 1 64 15 35 45, Fax : +33 1 64 15 35 62, Email : sulem@cermes.enpc.fr.

## Introduction

Hardness tests on rocks are currently used in rock mechanics and in rock engineering practice in order to provide a convenient and rapid characterisation of a rock. (see for example the comprehensive review of Atkinson [1]). Various quantitative measures of hardness depending on the particular test being employed are used empirically to characterise the rock drilling efficiency. For static indentation tests, an indentor is forced into a rock surface as for example the so-called 'stamp-test' [2] where a rigid circular indentor is used. Hardness is empirically related to the modulus of elasticity, the material yield stress, the fracture toughness, the material ductility or brittleness, the resistance to scratching, the surface energy. Moreover, apart from providing empirical applications in quality control, a more fundamental understanding of the indentation test in terms of actual deformation mechanisms has been developed in order to measure intrinsic fracture parameters of brittle solids [3-5]. This requires a detailed analysis of the actual indentation stress field which depends upon the nature of the contact zone and the size of this contact zone as compared to the characteristic size of the microstructure of the indented rock (e.g. grain size). The correlation between indentation hardness and grain size has been studied by Brace [6]. It was shown that the indentation hardness is proportional to $d^{-m}$ where d is the maximum grain diameter and m is a constant of the order of 0.5 for hard rocks and 0.3 for carbonates (see also Jaeger [7]).

Failure criteria for rocks usually involve only stresses and are thus suited primarily for homogeneous states of stresses. Since in rock mechanics highly inhomogeneous stresses may occur, it is possible that stress-gradients have some effects of failure mechanism. As mentioned by Mindlin [8], the apparent strength of rock-type materials is affected by strain gradient. It is observed that brittle failure and the onset of static yielding in the presence of stress concentration occur at higher loads than might be expected on the basis of stress concentration factors calculated from the theory of elasticity. In general, increasing strain gradients appear to make some materials stronger and to a degree that depends upon grain size. If the size of the indentor is comparable to the internal length of rock the scale effects cannot be neglected and the load-indentation curve must be re-interpreted in order to extract intrinsic parameters of the tested material.

Then it appears necessary to resort to continuum models with microstructure to describe correctly the deformation process at small scale. These generalized continua usually contain additional kinematical degrees of freedom (Cosserat continuum) and/or higher deformation gradients (higher grade continuum). Rotation gradients and higher velocity gradients introduce a material length scale into the problem, which allow to assess the effect of scale (e.g. Vardoulakis and Sulem [9]).

In this paper, the response of a Cosserat elasto-plastic half-space under indentation is studied using a large-strain finite element analysis with Cosserat structure [10]. The aim of this study is to show quantitatively that scale effect can be significant when the size of the indenting tool is comparable to the grain size of the rock. For granular rock, it has been shown in previous works (e.g. Vardoulakis and Sulem [9]) that the Cosserat theory is well suited to account for the influence of the microstructural response of the rock on the macroscopic behaviour.

The mechanisms of deformation and failure of the rock under the indentor are complex. In particular grain crushing may be important when the size of the indentor is rather big as compared to the grain size of the rock. However we are interested here in micro-indentation tests and in that case it is observed in micrographs of indented rock surface, that the dominant micro-mechanisms that control the deformation process is grain rotation and grain sliding without significant grain breakage (Fig.1). Therefor we shall consider in the numerical examples presented in this paper, a simple Mohr-Coulomb elastic perfectly plastic constitutive model with a Cosserat microstructure. Considering a perfectly plastic model, no mesh dependence in relation strain localisation and ill-posed mathematical problems encountered with softening behaviour is expected.

## A 2D-Cosserat elasto-plastic model

### *Kinematics and statics*

In a two-dimensional Cosserat continuum each material point has two translational degree of freedom ($u_1$ ,$u_2$) and one rotational degree of freedom $\omega^c$. The index c is used to distinguish the Cosserat rotation from the rotation

$$\omega = \frac{1}{2}\left(u_{2,1} - u_{1,2}\right) \quad ; \quad (.)_{,i} = \frac{\partial (.)}{\partial x_i} \quad i = 1,2 \tag{1}$$

For the formulation of the constitutive relationships we need deformation measures which are invariant with respect to rigid body motions which are the conventional strain tensor

$$\varepsilon_{ij} = \frac{1}{2}\left(u_{i,j} + u_{j,i}\right) \tag{2}$$

the relative rotation

$$\omega^{rel} = \omega - \omega^{c} \tag{3}$$

and the gradient of the Cosserat rotation which is called the curvature of the deformation

$$\kappa_i = \partial\omega^{c} / \partial x_i \tag{4}$$

It is usual to combine equations (2) and (3) to a single, tensorial deformation measure

$$\gamma_{11} = \partial u_1 / \partial x_1 \quad ; \quad \gamma_{12} = \partial u_1 / \partial x_2 + \omega^c$$
$$\gamma_{22} = \partial u_2 / \partial x_2 \quad ; \quad \gamma_{21} = \partial u_2 / \partial x_1 - \omega^c \tag{5}$$

The six deformation quantities (equations 4 and 5) are conjugate in energy to six stress quantities. First we have the four components of the non symmetric stress tensor $\sigma_{ij}$ which is conjugate to the non symmetric deformation tensor $\gamma_{ij}$ and second we have two couple stresses (moment per unit area) $m_1$ and $m_2$, which are conjugate to the two curvatures $\kappa_1$ and $\kappa_2$.

*Elastic strains*

The stress-strain relationships for a 2D isotropic Cosserat continuum are

$$\begin{aligned}
\gamma_{11} &= \frac{1}{2G}\left((1-\nu)\sigma_{11} - \nu\sigma_{22}\right) \\
\gamma_{22} &= \frac{1}{2G}\left(-\nu\sigma_{11} + (1-\nu)\sigma_{22}\right) \\
\gamma_{12} &= \frac{1}{4G}\left(\sigma_{12} + \sigma_{21}\right) + \frac{1}{4G^c}\left(\sigma_{12} - \sigma_{21}\right) \\
\gamma_{21} &= \frac{1}{4G}\left(\sigma_{12} + \sigma_{21}\right) - \frac{1}{4G^c}\left(\sigma_{12} - \sigma_{21}\right) \\
\kappa_1 &= \frac{1}{M} m_1 \quad , \quad \kappa_2 = \frac{1}{M} m_2
\end{aligned} \tag{6}$$

In equation (6) $G$ and $\nu$ are the classical elastic shear modulus and Poisson's ratio respectively. The additional Cosserat shear modulus $G^c$ links the antisymmetric part of the deformation tensor to the antisymmetric part of the stress tensor and couple-stresses and curvatures are linked through a bending modulus $M$, which has the dimension of a force. Thus in 2D Cosserat elasticity the problem is governed by four material constants. An internal material length for bending appears as

$$\ell = \sqrt{M / G} \tag{7}$$

*Plastic strains*

For more convenient representation we introduce first the pseudo vectors

$$\begin{aligned} \boldsymbol{s} &= \left\{\sigma_{11}, \sigma_{22}, \sigma_{12}, \sigma_{21}, m_1, m_2\right\}^T \\ \boldsymbol{e} &= \left\{\gamma_{11}, \gamma_{22}, \gamma_{12}, \gamma_{21}, \kappa_1, \kappa_2\right\}^T \end{aligned} \tag{8}$$

the usual decomposition between elastic and plastic strains is used

$$\dot{\boldsymbol{e}} = \dot{\boldsymbol{e}}^e + \dot{\boldsymbol{e}}^p \tag{9}$$

A 2D flow theory of plasticity for granular media with Cosserat microstructure can be derived by keeping the same definitions for the yield surface and the plastic potential as in the classical theory and by generalising appropriately the stress and strain invariants involved in these definitions (Vardoulakis and Sulem [9], Lippmann [11]). The following generalised stress invariants are utilised

$$\sigma = \sigma_{kk} / 2$$
$$\tau = \sqrt{h_1 s_{ij} s_{ij} + h_2 s_{ij} s_{ji} + h_3 \frac{m_k m_k}{\ell^2}} \quad ; \quad h_1 + h_2 = 1/2 \tag{10}$$

with

$$s_{ij} = \sigma_{ij} - \sigma \delta_{ij} \tag{11}$$

We distinguish among a static and a kinematical plasticity model [9]. It can be shown that a kinematical model which is based on the micromechanical definition of interparticle slip results in the following set of weighting factors,

$$h_1 = \frac{3}{8} \quad ; \quad h_2 = \frac{1}{8} \quad ; \quad h_3 = \frac{1}{4} \tag{12}$$

A static model which is more appropriate for granular rocks is based on the micromechanical definition of interparticle shear and is given by

$$h_1 = \frac{3}{4} \quad ; \quad h_2 = -\frac{1}{4} \quad ; \quad h_3 = 1 \qquad (13)$$

The elastic Cosserat parameters can be defined from micromechanical considerations [12-13]:

$$G^c = \frac{1}{2(h_1 - h_2)} G; \quad M = GR_g^2 / h_3 \qquad (14)$$

where $R_g$ is the grain diameter.

Consequently:

for the statical model $G_c = G/2$ ; $\ell = R_g$

for the kinematical model $G_c = 2G$ ; $\ell = R_g/\sqrt{2}$

By analogy to the classical flow theory of plasticity, a Mohr-Coulomb yield surface and plastic potential are defined as

$$\begin{aligned} F &= \tau - \sin\phi(q - \sigma) = 0 \\ Q &= \tau + \sigma \sin\psi \end{aligned} \qquad (15)$$

where $\sigma$ and $\tau$ are the generalised invariants of the plane Cosserat continuum defined above, $\phi$ and $\psi$ are the mobilised friction and dilatancy angles respectively and $q$ is the tension cut-off. The mobilised cohesion of the rock is $c = q\cot\phi$.

In flow theory of plasticity, the flow rule states that the plastic strain increments are proportional to a given vector which is normal to the plastic potential surface

$$\dot{\boldsymbol{e}}^{p} = \dot{\lambda}\frac{\partial Q}{\partial \boldsymbol{s}} \tag{16}$$

with

$$\frac{\partial Q}{\partial \mathbf{s}} = \begin{bmatrix} (h_1 + h_2)\frac{s_{11}}{\tau} + \frac{1}{2}\sin\psi \\ (h_1 + h_2)\frac{s_{22}}{\tau} + \frac{1}{2}\sin\psi \\ h_1\frac{s_{12}}{\tau} + h_2\frac{s_{21}}{\tau} \\ h_2\frac{s_{12}}{\tau} + h_1\frac{s_{21}}{\tau} \\ h_3\frac{m_1}{\tau} \\ h_3\frac{m_2}{\tau} \end{bmatrix} \tag{17}$$

Similarly the normal to the yield surface in generalised stress space is obtained by replacing in equation (15) the dilatancy angle by the friction angle.

*Incremental elasto-plastic constitutive equations*

Elasticity relations (6) are written under the general form

$$\dot{\boldsymbol{s}} = \boldsymbol{C}^{e}\dot{\boldsymbol{e}}^{e} \tag{18}$$

where $\boldsymbol{C}^e$ is the elastic stiffness tensor.

In equation (15) the plastic multipliers where $\dot{\lambda}$ are eliminated by using as in classical plasticity the consistency conditions

$$F = 0 \quad \text{and} \quad \dot{F} = \left(\frac{\partial F}{\partial \boldsymbol{s}}\right)^T \dot{\boldsymbol{s}} + \left(\frac{\partial F}{\partial \boldsymbol{e}^p}\right)^T \dot{\lambda} \frac{\partial Q}{\partial \boldsymbol{s}} = 0 \tag{19}$$

If we assume no hardening we obtain the following explicit form of the incremental constitutive equations

$$\dot{s}_i = C_{im}\dot{e}_m = C^e_{ik}\left[\delta_{km} - \frac{\dfrac{\partial Q}{\partial s_k} C^e_{mn} \dfrac{\partial F}{\partial s_n}}{\dfrac{\partial F}{\partial s_r} C^e_{rs} \dfrac{\partial Q}{\partial s_s}}\right]\dot{e}_m \quad \text{or} \quad \dot{\boldsymbol{s}} = \boldsymbol{C}\dot{\boldsymbol{e}} \tag{20}$$

**Finite element formulation**

For general boundary value problems numerical methods are used. The finite element method is a well established tool for these purposes. Finite element analyses for a Cosserat continuum has been presented by several authors [14-17]. In the following we present the most important features of the extension of the method to a Cosserat continuum.

We define general displacement and traction pseudo vectors as

$$\boldsymbol{v} = \left\{ u_1, u_{2,}, \ell\omega^c \right\}^T \quad ; \quad \boldsymbol{f} = \left\{ t_1, t_2, \frac{m_3}{\ell} \right\}^T \tag{21}$$

by means of which the virtual work principal can be written as

$$\int_B \boldsymbol{s}^{\boldsymbol{T}} \delta \boldsymbol{e} dV = \int_{\partial B} \boldsymbol{f}^{\boldsymbol{T}} \delta \boldsymbol{v} dA \tag{22}$$

Above equation (21) looks formally the same as for the classical continuum. Using the constitutive relationships (19) the incremental form of (21) is

$$\int_B \Delta \boldsymbol{e}^{\boldsymbol{T}} \boldsymbol{C}^T \delta \boldsymbol{e} dV = \int_{\partial B} \boldsymbol{f}^{\boldsymbol{T}} \delta \boldsymbol{v} dA - \int_B \boldsymbol{s}_t^T \delta \boldsymbol{e} dV \tag{23}$$

Essentially the finite element method consists of specifying an assumed distribution of the displacements and rotations within the domain $B^e$ of a finite element. This can be written as

$$\boldsymbol{v} = \phi^M \boldsymbol{v}^M \quad ; \delta \boldsymbol{v} = \phi^M \delta \boldsymbol{v}^M \quad M = 1,2 \ldots . M^e \tag{24}$$

where $\phi^M$ are the so-called shape functions, $M$ is the number of nodal points and $M^e$ the total number of nodal points of each element. When dealing with Cosserat medium the question of the order of interpolation of the shape functions arises. Since the same order of derivatives of translational displacements and the rotations is involved in the governing equations, the shape functions of the displacement should be one order higher that those of the Cosserat rotation. This is however not considered at the moment in our finite element code where the linear interpolation for displacements and rotations is used for 4-noded element.

The relation between the deformation vector $\boldsymbol{e}$ and the nodal variables is written as :

$$\boldsymbol{e} = \boldsymbol{B}^M \boldsymbol{v}^M \quad ; \quad \left[\boldsymbol{B}^M\right] = \begin{bmatrix} \phi^M_{,1} & 0 & 0 \\ 0 & \phi^M_{,2} & 0 \\ \phi^M_{,2} & 0 & \phi^M \\ 0 & \phi^M_{,1} & -\phi^M \\ 0 & 0 & \phi^M_{,1} \\ 0 & 0 & \phi^M_{,2} \end{bmatrix} \tag{25}$$

Inserting equation (24), equation (22) becomes

$$\left(\delta \boldsymbol{v}^M\right)^T \boldsymbol{K}^{MN} \Delta \boldsymbol{v}^N = \left(\delta \boldsymbol{v}^M\right)^T \left(\boldsymbol{F}^M_{ext} - \boldsymbol{F}^M_{\text{int}}\right) \tag{26}$$

where

$$\boldsymbol{K}^{\boldsymbol{MN}} = \int_{B^e} \left(\boldsymbol{B}^M\right)^T \boldsymbol{C}\boldsymbol{B}^{\boldsymbol{N}} dV \tag{27}$$

is the element tangent stiffness matrix,

$$\left[\boldsymbol{F}_{\boldsymbol{ext}}^{\boldsymbol{M}}\right] = \begin{bmatrix} \displaystyle\int_{\partial_1 B^e} \boldsymbol{t}\phi^{\boldsymbol{M}} dA \\ \displaystyle\int_{\partial_2 B^e} \frac{m}{\ell}\phi^{\boldsymbol{M}} dA \end{bmatrix} \tag{28}$$

is the generalised external load vector and

$$\left[\boldsymbol{F}_{\text{int}}^{M}\right] = \int_{B^e} \left(\boldsymbol{B}^M\right)^T \boldsymbol{s}_{\boldsymbol{t}} dV \tag{29}$$

is the generalised initial stress vector.

The above formulation can be extended to large strain deformation as proposed by Adhikary et al [18]. This is done on one side by updating of the finite element mesh at each step of loading and on the other side by considering in the expression of the tangent stiffness matrix an additional term which results from the consideration of geometric non-linearities [19]. Equation (26) is thus modified as follows:

$$\boldsymbol{K}^{MN} = \int_{B^e} \left(\boldsymbol{B}^M\right)^T \boldsymbol{C}\boldsymbol{B}^N dV + \boldsymbol{K}_{Geom}^{NM} \tag{30}$$

where

$$\boldsymbol{K}_{Geom}^{NM} = \int_{b^e} \begin{vmatrix} 0 & 0 & \left(-\sigma_{12}\phi_{N,1} - \sigma_{22}\phi_{N,2}\right)\phi_M \\ 0 & 0 & \left(\sigma_{11}\phi_{N,1} + \sigma_{21}\phi_{N,2}\right)\phi_M \\ \left(-\sigma_{12}\phi_{M,1} - \sigma_{22}\phi_{M,2}\right)\phi_N & \left(\sigma_{11}\phi_{M,1} + \sigma_{21}\phi_{M,2}\right)\phi_N & \left(-\sigma_{11} - \sigma_{22}\right)\phi_M\phi_N \end{vmatrix} dV \tag{31}$$

Notice that equation (30) is identical to equation (45) in Adhikary et al [18] except a small difference in sign for the term $(-\sigma_{11}-\sigma_{22})\phi_M\phi_N$.

## Numerical analysis of the indentation test

### *Position of the problem and material parameters*

An axisymmetric finite element simulation of an indentation experiment on a elastic-perfectly plastic granular rock has been performed in order to study scale effect. As experimental data on such scale effect are not available at the moment for rocks, the aim of this study is to evaluate the influence of the various parameters and to compare the results of the simulation for a Cosserat continuum and for a classical Cauchy medium.

The numerical analysis is using the above bidimensional large-strain Cosserat theory adapted to axisymmetric problems. The effect of punching of a flat, circular and rigid indentor of size $2a$ is simulated by applying constant displacement at the corresponding nodes of the finite element mesh. This boundary condition is considered in our finite element code by putting elastic springs with infinite stiffness defined in the vertical ($x_2$) direction and given displacements at the considered nodes. To simulate the behaviour of the interface between the rock and the punching tool in the horizontal ($x_1$) direction, two cases were considered herein:

a) the case of allowed horizontal ($x_1$) displacements (perfect sliding) of the nodes at the interface between rock and the punching tool is shown in Fig. 2.a. and
b) the case of zero horizontal ($x_1$) displacements (perfect adherence) of the nodes at the interface between rock and the punching tool is shown in Fig. 2.b, where a second set of springs with infinite stiffness and zero imposed displacements in the horizontal ($x_1$) direction is added to the model.

The easiest way to simulate the perfect adherence condition between the rock and the punching tool is to use infinite-stiffness springs, since they allow to impose null displacements to the model by avoiding numerical contamination. The two extreme cases of free horizontal displacement (perfectly sliding contact) and totally restrained horizontal displacement (perfectly adherent contact) at the rock-tool interface is then studied without the need of introducing an interface frictional model. On the other hand the higher order boundary condition (in terms of imposed Cosserat rotation or imposed couple stress) introduces a 'boundary length' and therefor it is possible in the frame of the present theory to take into account the roughness of the tool. At the present time without experimental data on the effect of this parameter it is assumed in the numerical simulation that no couple stress is imposed on the boundaries of the model.

From the computed normal stress on the loaded surface it is then possible to calculate the corresponding punching force. The following set of material constants has been used for a sandstone:

Young modulus E = 18 000 MPa, Poisson's ratio $\nu = 0.11$, friction angle $\phi = 38°$, uniaxial compression strength UCS = 46 MPa. For the additional Cosserat parameters namely the Cosserat shear modulus $G_c$ and the internal length we shall distinguish between the kinematical and the statical Cosserat model as presented in the previous section.

*Spatial discretisation and boundary conditions*

For symmetry reasons, only half of the domain is discretised as shown on Fig. 3. Zero horizontal ($x_1$) displacements and Cosserat rotations are maintained at the axis of symmetry (AC) as well as on the boundary (BD). Zero vertical ($x_2$) displacement and Cosserat rotation are maintained on boundary (DC). It is also assumed that on (AB) no couple-force is applied which simulates the effect of a perfectly smooth tool at the interface and of a free surface elsewhere.

Two different mesh sizes are tested. For the first one M1 (Figure 4a) which is rather coarse the size of the square elements under the indentor is $a/4$. For the second one M2 which is finer the corresponding element size is $a/7$ (Figure 4b).

*Elastic response*

Assuming no friction at the interface between the indentor and the rock, indentation of a linear elastic half space is simulated for various values of the scale parameter $\ell/2a$ assuming a statical elastic Cosserat model as shown on Fig. 5. The response is linear and the slope of the indentation curve is increasing with the scale as $\ell/2a$ increases which means that the apparent rigidity of the rock will be bigger for small indentors than for large ones as compared to the internal length (grain size) of the material. If the size of the indentor is comparable to the internal length (grain size) of the material the apparent rigidity of the rock is bigger than in the case of a large indentor as compared to the material length. This is summarised on Table 1 for the two different mesh sizes.

Usual indentors for rocks have a size comprised between 0.5 to 2 mm. If the size of the indentor is comparable to the grain size of the material the apparent rigidity is overestimated of about 18%.

These results can be compared to the analytic solution of the response of a classical (Cauchy) elastic half-space with uniform normal displacement applied on a circular region [20] where the slope of the indentation curve is $\frac{2a}{1-\nu^2}E = 18220\mathrm{N/mm}$. Table 1 shows also that the results obtained with the two different mesh sizes are very close.

The effect of the conditions at the interface between the indentor and the rock are studied by considering the two extreme cases of perfect sliding and perfect adherence. Scale effect is shown on Fig. 6 where the slope of the indentation response for a Cosserat elastic continuum and a classical Cauchy elastic continuum is plotted as a function of the scale factor $\ell/2a$. This figure shows that the scale effect is still more pronounced in case of frictionless interface (18%) that in case of perfectly adherent one (12%).

*Scale effect for an elasto-plastic Cosserat medium*

The aim of the computations presented in this section is to show that the Cosserat model is able to capture a scale effect in terms of maximum indentation force as a function of the scale parameter $\ell / 2a$. We therefor consider here a simple elastic perfectly plastic Cosserat model which is an acceptable assumption for granular rock.

a) Perfectly sliding rock-tool interface

A simulated indentation curve for $\ell / 2a = 0.5$ and a statical model is shown on Fig. 7 assuming perfect sliding of the rock-tool interface. The responses obtained for the coarse mesh M1 and for the more refined one M2 are shown on this figure. Obviously the two simulations lead to close results which gives confidence in the fact that our results are not mesh-dependent. The corresponding deformed mesh with plastic zones indicated in dark colour for an imposed displacement of 0.1 mm on the loaded zone is shown for both models on Fig. 8.

On Fig. 9, the simulated indentation curves for various values of the scale parameter are shown for a statical Cosserat model (Fig. 9a) and for a kinematical one (Fig. 9b). These figures show that the maximum load is increasing with increasing values of the scale parameter. The softening part of the curve is purely structural and is due to stress redistribution when the maximum strength is reached. It is observed that this structural softening is less important for higher values of the scale parameter due to arching effect of the Cosserat microstructure. The scale effect is shown on Fig. 10 with comparison between the results obtained assuming a statical Cosserat model and a kinematical Cosserat. The maximum force for an indentation depth 0.2mm ($a$=0.5mm) is plotted versus the ratio $\ell / 2a$. This curve shows that for a classical (Cauchy) continuum ($\ell = 0$) the apparent strength is smaller as the one obtained for a Cosserat continuum for which the internal length is comparable to the indentor size ($\ell / 2a = 0.5$). This scale effect can reach 15% for the statical model and 50% for the kinematical model. For the same loading force (F = 240N) the deformed meshes obtained with a statical model for two different internal lengths are compared on Fig.11. For $\ell / 2a = 0.01$ (Fig. 10a) the displacement under the indentor is 0.03mm whereas for $\ell / 2a = 0.5$ (Fig. 10b) the displacement under the indentor is 0.016 mm i.e. about twice smaller.

b) Effect of frictional rock-tool interface

The effect of the interface conditions on the scale effect is studied by comparing the above results to the one obtained in the extreme case of perfect adherence between the tool and the rock. The simulated indentation curves for a statical Cosserat model and a kinematical model assuming zero displacement in $x_1$ direction at the interface nodes are shown on Fig. 12a and Fig. 12b respectively, for various values of the scale factor. These graphs which are to be compared with the one of Fig.9a,b (for which perfectly sliding interface is assumed) show that the necessary force to indent the rock of a given displacement is higher in case of adherent interface than for sliding interface. It shows also that the loading force is monotonous without structural softening as obtained in the other case. The scale effect for the two extreme interface conditions is shown on Fig.13a,b. This graph shows that the scale effect is of the same order (15% for the statical model and 50% for the kinematical model) for both interface conditions. This shows that interface friction is not influencing significantly scale effect.

## Conclusions

Hardness tests are currently used in rock mechanics to characterise rock mechanical properties and rock drilling efficiency. However it is observed that brittle failure is influenced by large strain gradients and that the onset of static yielding in the presence of stress concentration occurs at higher loads than might be expected from classical continuum theories. Although based on simple constitutive assumptions and geometrical configuration the above analysis gives an example of microstructural effects in the presence of stress concentration. Using finite element numerical simulations, it is shown that for a material with Cosserat microstructure, the apparent strength and rigidity increase as the size of the indentor decreases. This scale effect for the strength can reach 15% for a statical model and 50% for a kinematical Cosserat model when the size of the indentor tool is comparable to the grain size of the rock. It is shown that this scale effect is not significantly affected by the interface condition at the rock tool interface. Such a scale effect has been observed experimentally for [21, 22]. In the lack of relevant quantitative experimental data for the scale effect in the case of rocks this analysis suggests that this effect may be of importance and has to be investigated further. In addition, indentation tests appear as an experimental tool for the testing and validation of continuum theories with microstructure and calibration of internal lengths parameters.

**References**

[1] Atkinson R.H. Hardness tests for rock characterisation. In: Hudson J, Editor. Comprehensive Rock Engineering, Pergamon Press 1993; vol. 1, chap. 5 p. 105-117

[2] Wijk G. The stamp test for rock drillability classification. Int. J. Rock Mech. and Min. Sci., 1989; 26(1): 37-44.

[3] Lawn B. and Wilshaw R. Indentation fracture: principles and applications. J. Mat. Sci., Chapman and Hall, 1975; 10: 1049-1081.

[4] Detournay E. and Defourny P. A phenomenological model for the drilling action of drag bits. Int. J. Rock Mech. And Min. Sci., 1992; 29(1): 13-23.

[5] Huang H., Damjanac B. and Detournay E. Numerical modelling of normal wedge indentation in rocks with lateral confinement. Int. J. Rock Mech. and Min. Sci. 1997; 34 (3-4), paper No. 064.

[6] Brace WF. Dependence of fracture strength of rocks on grain size. Penn. State Mineral Ind. Expl. Sta. Bulletin, 1961; 76: 99-103.

[7] Jaeger JC. Brittle failure of rocks, in: Failure and Breakage of Rocks. Proc. 8$^{th}$ U.S. Symposium on rock mechanics, 1967;3-57.

[8] Mindlin RD. The influence of couple stresses on stress concentrations. Experimental Mech. 1963; 3, 1-7.

[9] Vardoulakis I. and Sulem J. Bifurcation analysis in geomechanics. Blackie Academic & Professional, 1995.

[10] Cerrolaza M., Sulem J. and El Bied A. A Cosserat non-linear finite element analysis software for blocky structures. Int. J. Adv. Eng. Software, 1999; 30: 69-83.

[11] Lippmann H. Cosserat plasticity and plastic spin. ASME Appl. Mech. Rev. 1995; 48(11): 753-762.

[12] Mühlhaus HB. and Vardoulakis I. The thickness of shear bands in granular materials. Géotechnique 1987; 37: 271-283.

[13] Sulem J. and Vardoulakis I. Bifurcation analysis of the triaxial test on rock specimens. A theoretical model for shape and size effect. Acta Mechanica 1990; 83: 195-212.

[14] Borst R. de and Sluys LJ. Localisation in a Cosserat continuum under static and dynamic loading conditions. Comp. Meth. Appl. Mech. Eng. 1991; 90: 805-827.

[15] Papanastasiou P. and Vardoulakis I. Numerical treatment of progressive localisation in relation to borehole stability. Int. J. Num. Anal. Meth. Geomech. 1992; 16: 389-424.

[16] Dai C., Mühlhaus HB., Duncan Fama M, Meek J. Finite element analysis of Cosserat theory for layered rock mass. Computers and Geotechnics 1993; 15: 145-162.

[17] Ehlers W. and Volk W. On shear band localisation phenomena of liquid-saturated granular elastoplastic porous solid material accounting for fluid

viscosity and micropolar solid rotations. Mech. Cohesive-Frictional Mat. 1997; 2: 301-320.

[18] Adhikary DP., Mühlhaus HB. and Dyskin AV. Modelling the large deformations in stratified media – the Cosserat continuum approach. Mechanics of Cohesive-Frictional Mat. 1999; 4: 195-213.

[19] Sulem J. and Mühlhaus HB. A continuum model for periodic two-dimensional block structures. Mechanics of Cohesive-Frictional Mat. 1997; 2: 31-46.

[20] Johnson KL. Contact mechanics, Cambridge University Press, 1985.

[21] Poole WJ., Ashby MF. and Fleck NA. Micro-hardness of annealed and work-hardened copper polycrystals. Scripta Materialia, , Elsevier Science Ldt 1996; 35(4): 559-564.

[22] Fleck, N.A. and J.W. Hutchinson. Strain gradient plasticity. Advances in Appl. Mech.1997; 33: 295-361.

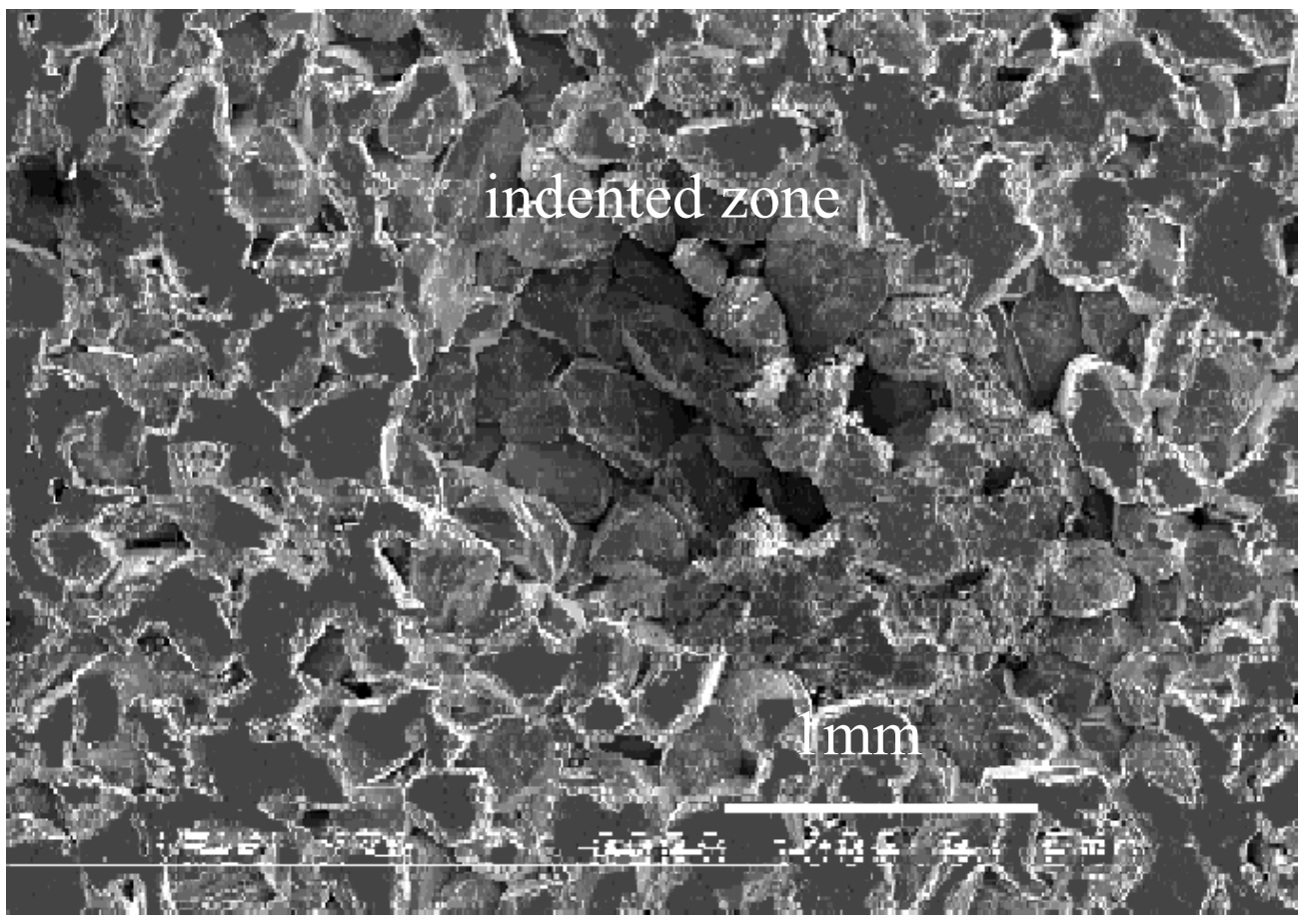

Figure 1: Fontainebleau sandstone indented with a flat indentor

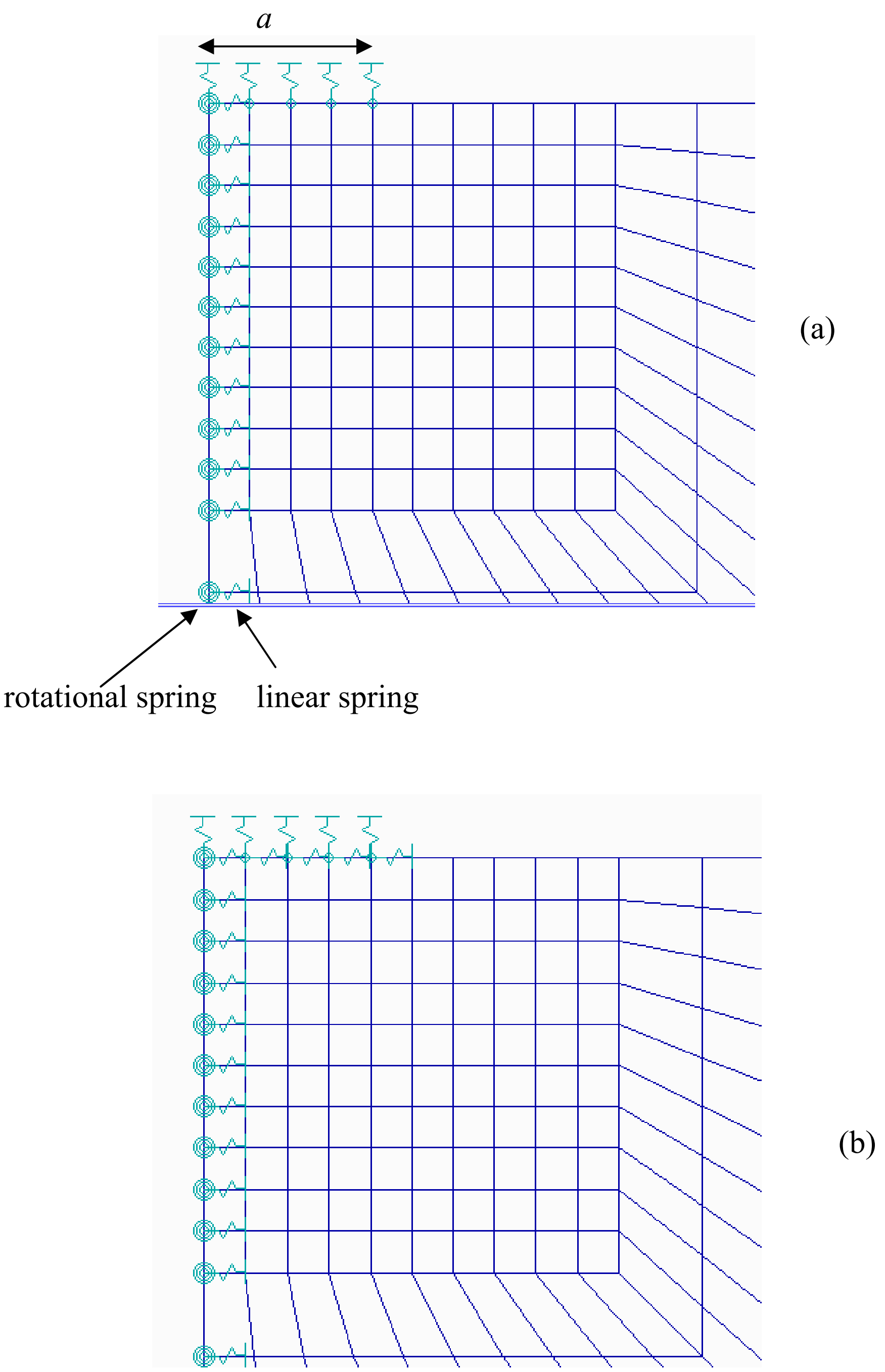

Figure 2: Modelling boundary conditions in terms of horizontal ($x_1$) displacements at the rock-punching tool interface
(a) Perfect sliding (allowed displacement $x$1-axis) ; (b) Perfect adherence (zero displacement along $x_1$-axis)

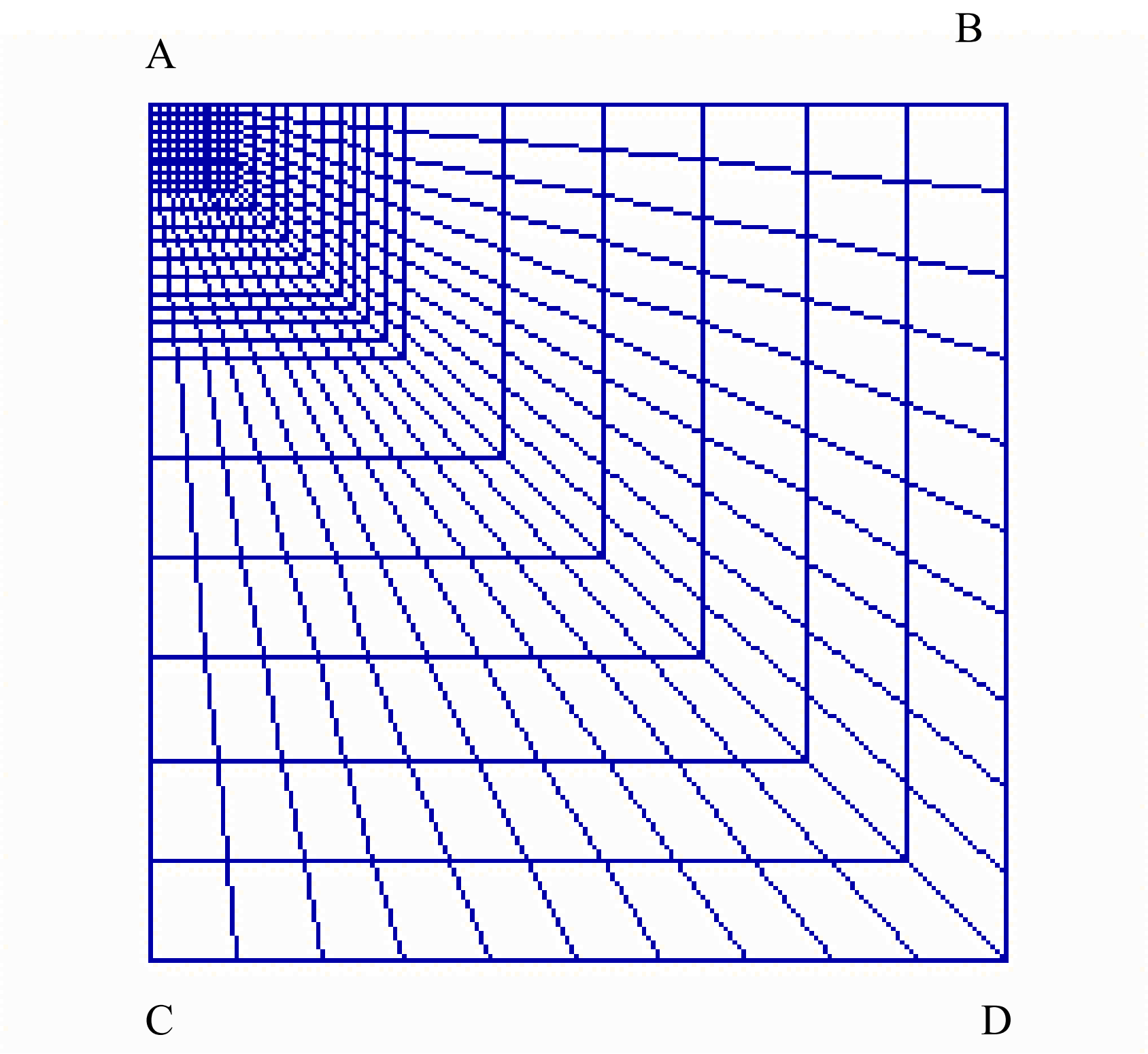

Figure 3 : The complete finite element model for mesh M1

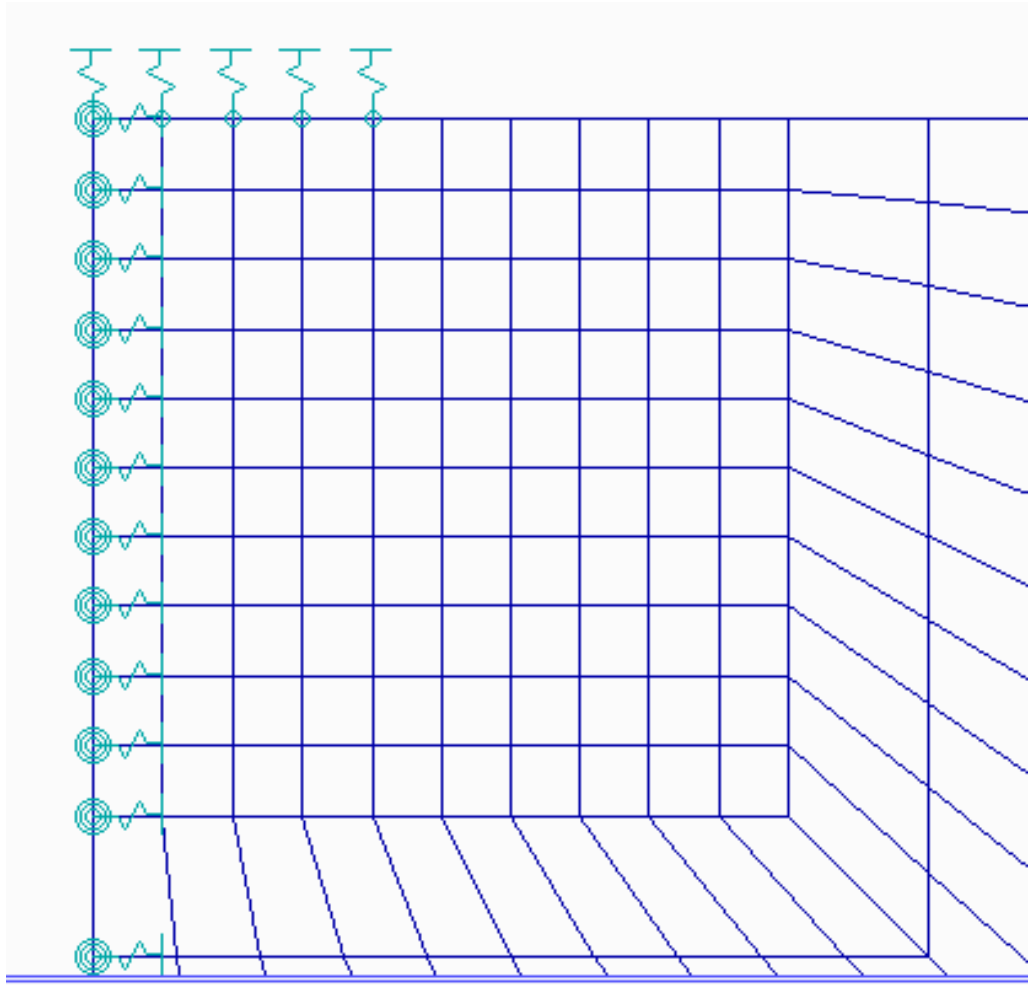

(a)

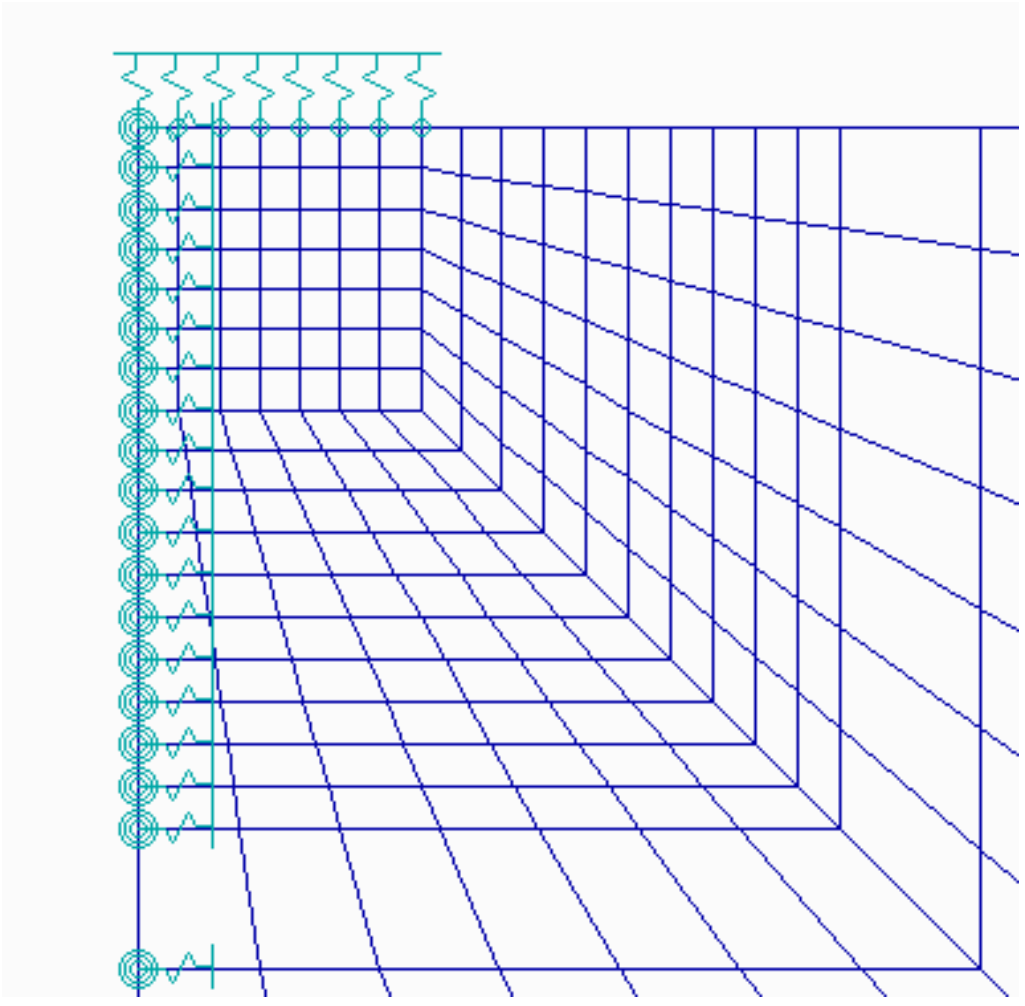

(b)

Figure 4 : The finite element mesh zoomed on the zone of loading: (a) model M1, (b) model M2

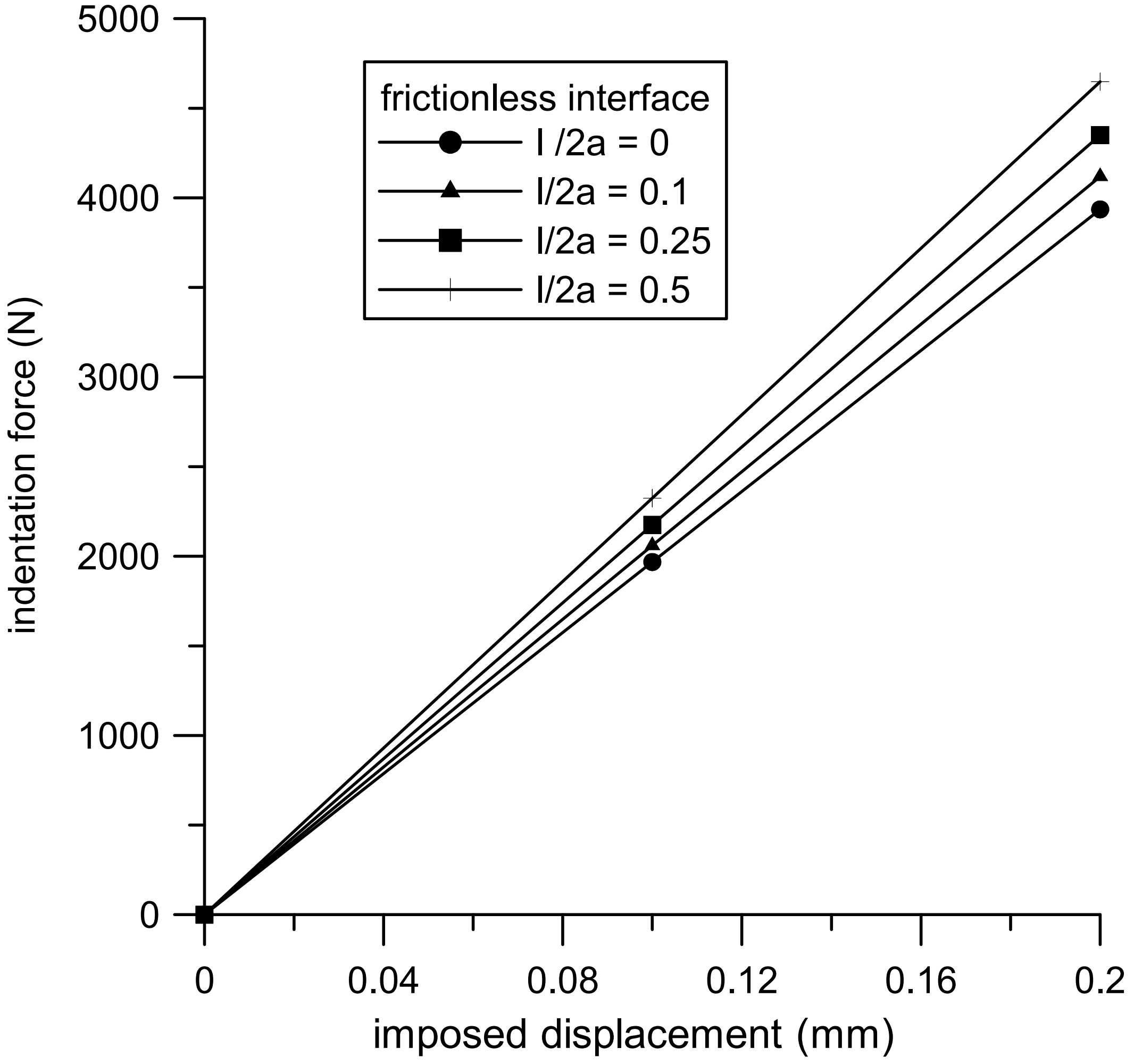

Figure 5: Elastic response for various values of the scale factor (statical Cosserat model), $a$=0.5mm

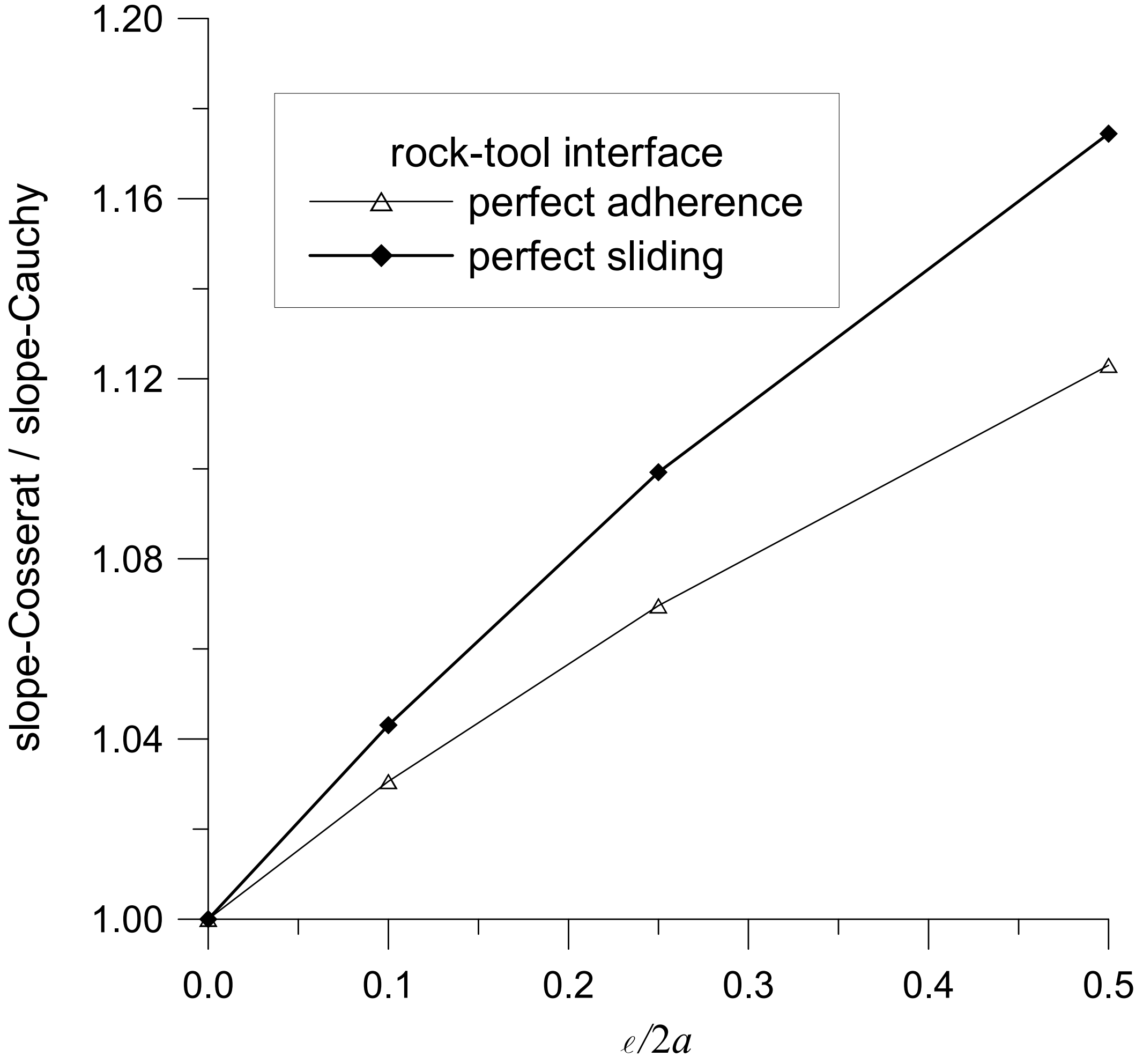

Figure 6: Effect of interface conditions on scale effect for an elastic Cosserat half-space

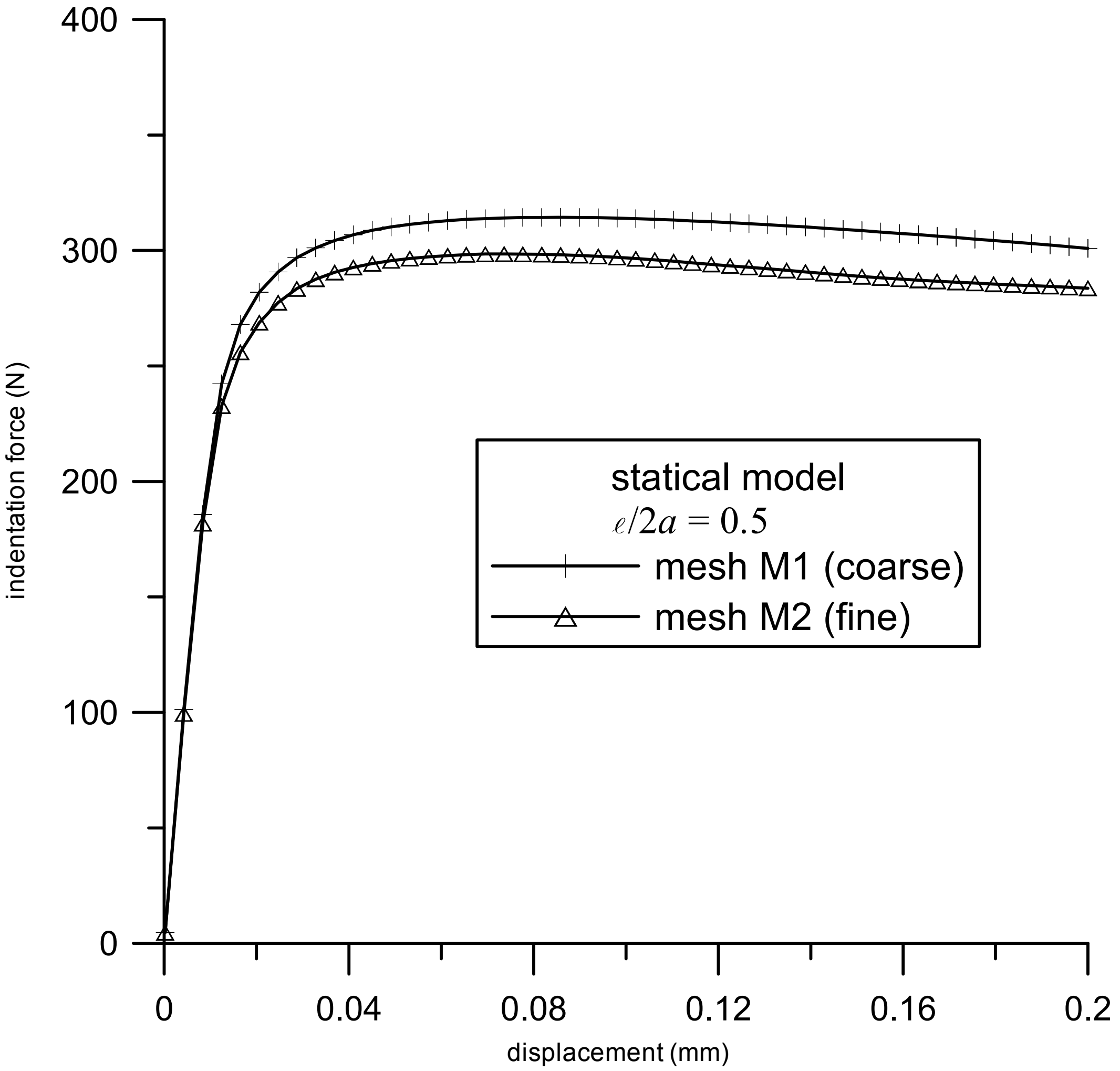

Figure 7: Simulation of an indentation experiment on an elasto-plastic Cosserat half-space for $\ell / 2a = 0.5$ with two different mesh sizes (statical model).

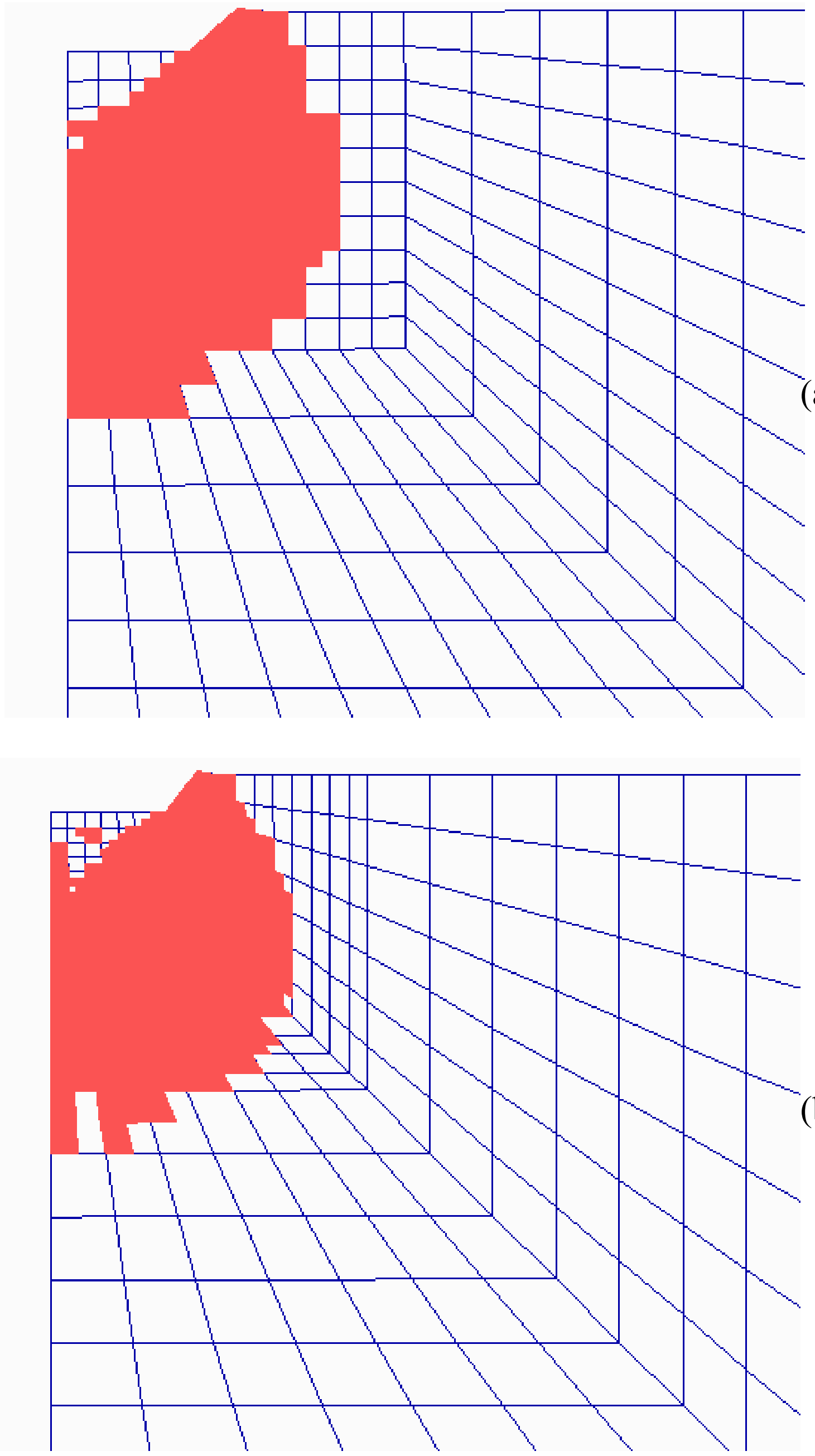

(a)

(b)

Figure 8: Deformed mesh with plastic zones for an imposed displacement of 0.1 mm on the boundary; (a) mesh M1, (b) mesh M2 (frictionless interface)

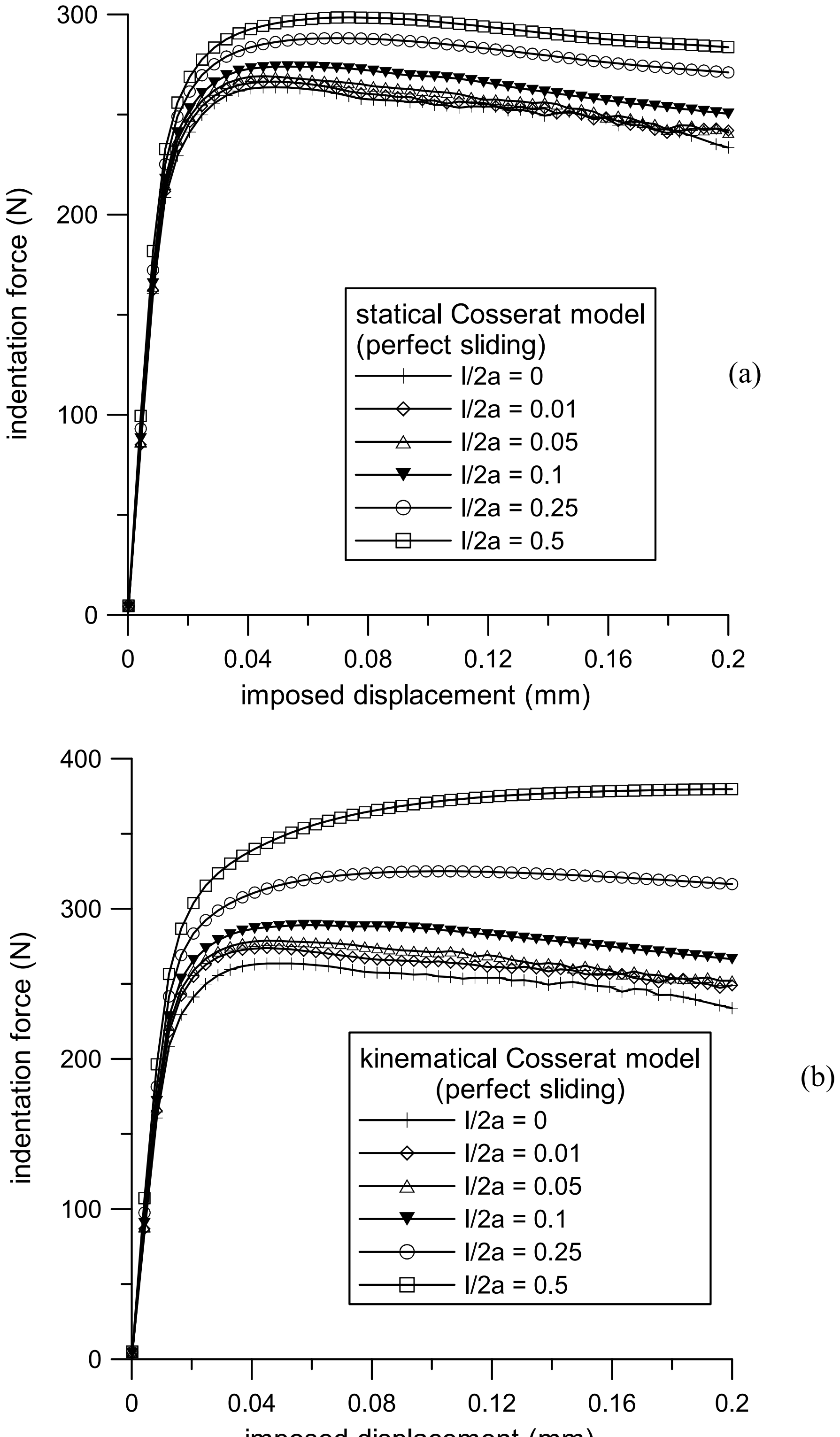

Figure 9 : Elastic-perfectly plastic Cosserat continuum :Computed indentation curves assuming perfect sliding conditions at rock tool interface (*a* = 5mm)
(a) statical model, (b) kinematical model

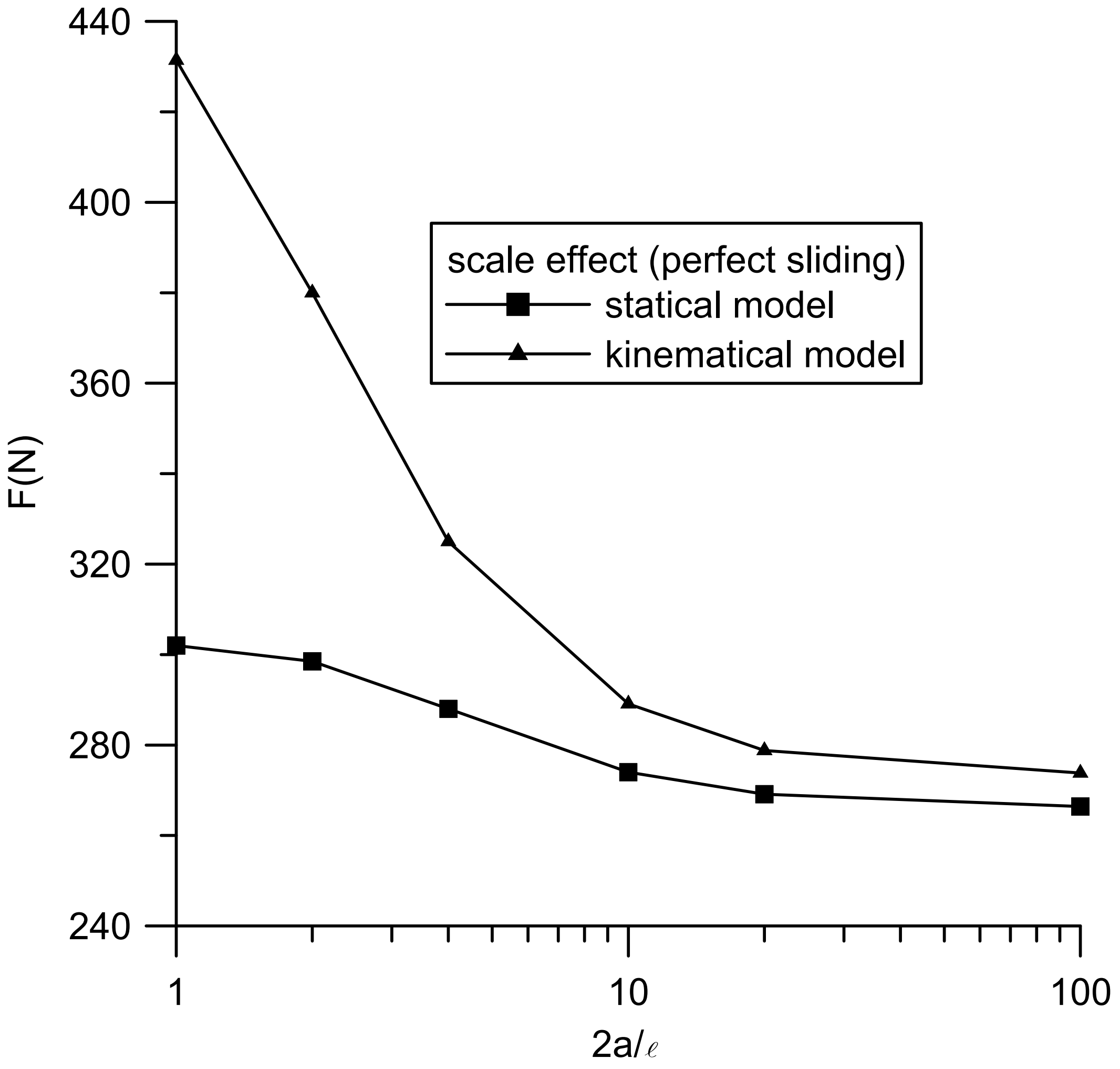

Figure 10: Scale effect for the maximum load of a Cosserat elastic-perfectly plastic rock under indentation (frictionless rock-tool interface)

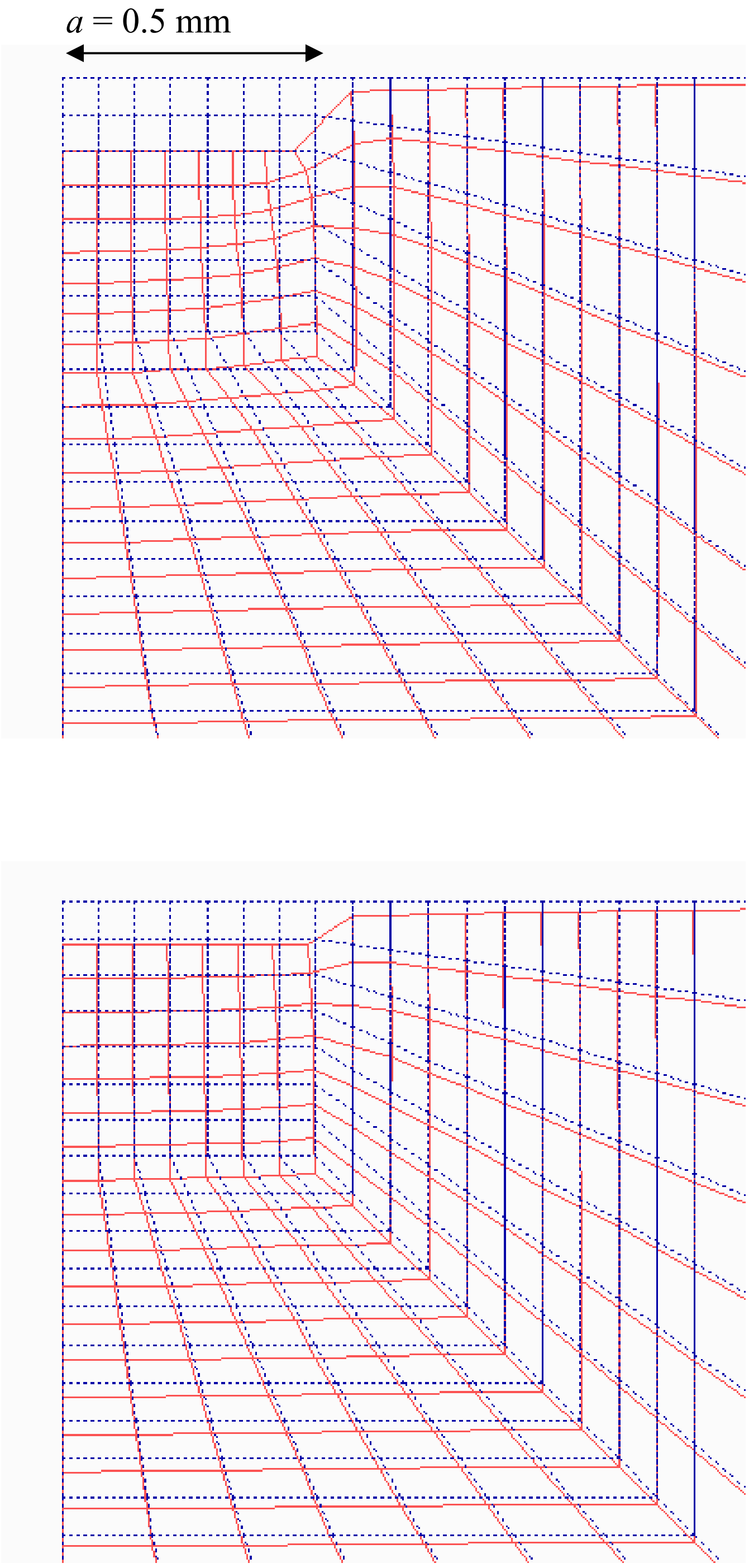

(a)

(b)

Figure 11: Deformed mesh for a loading force of 240 N (statical model, frictionless interface)
(a) $\ell / 2a = 0.01$, (b) $\ell / 2a = 0.5$

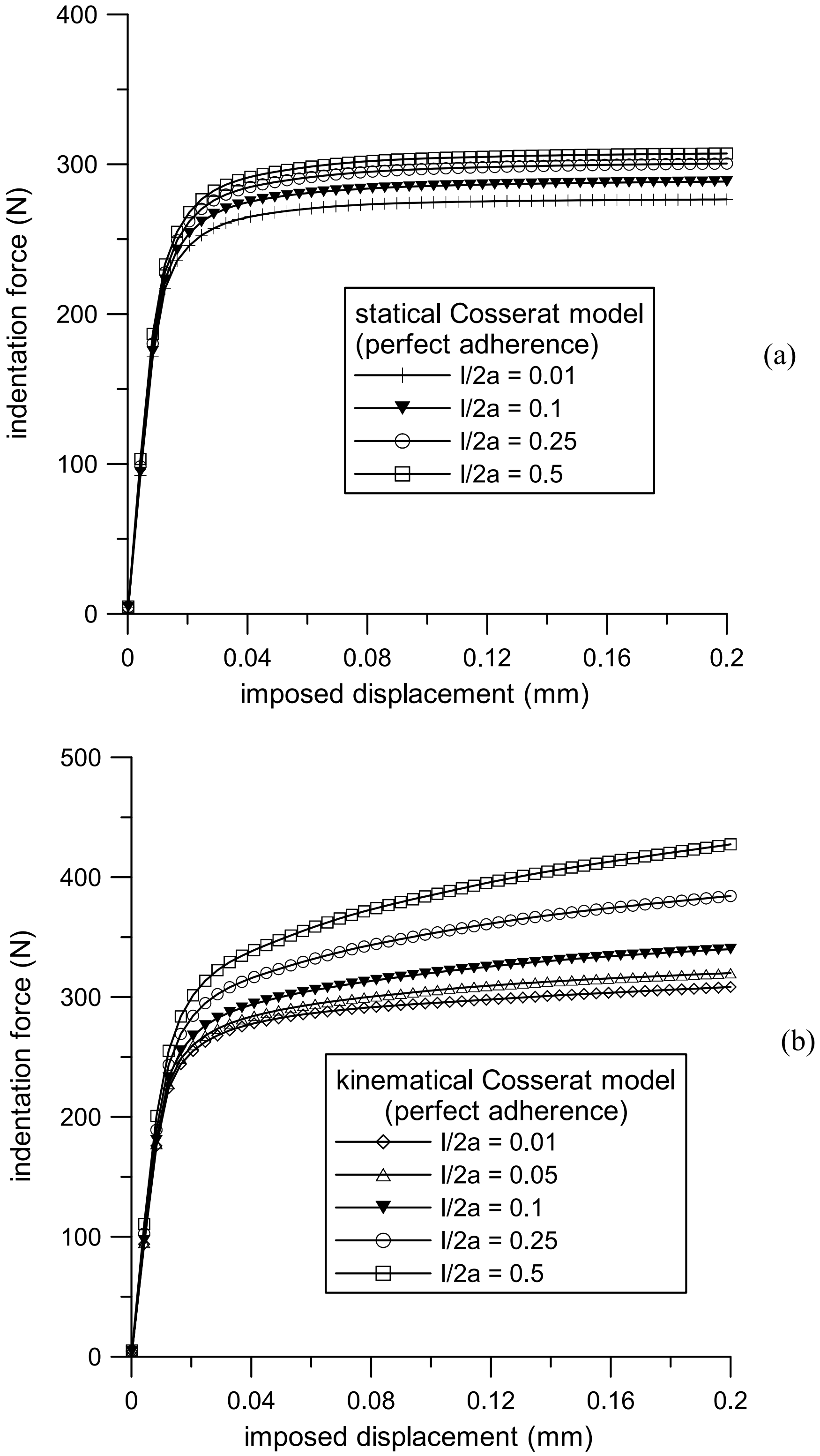

Fig. 12 : Elastic-perfectly plastic Cosserat continuum :Computed indentation curves assuming perfect adherence at rock tool interface ($a$ = 5mm)
(a) statical model, (b) kinematical model

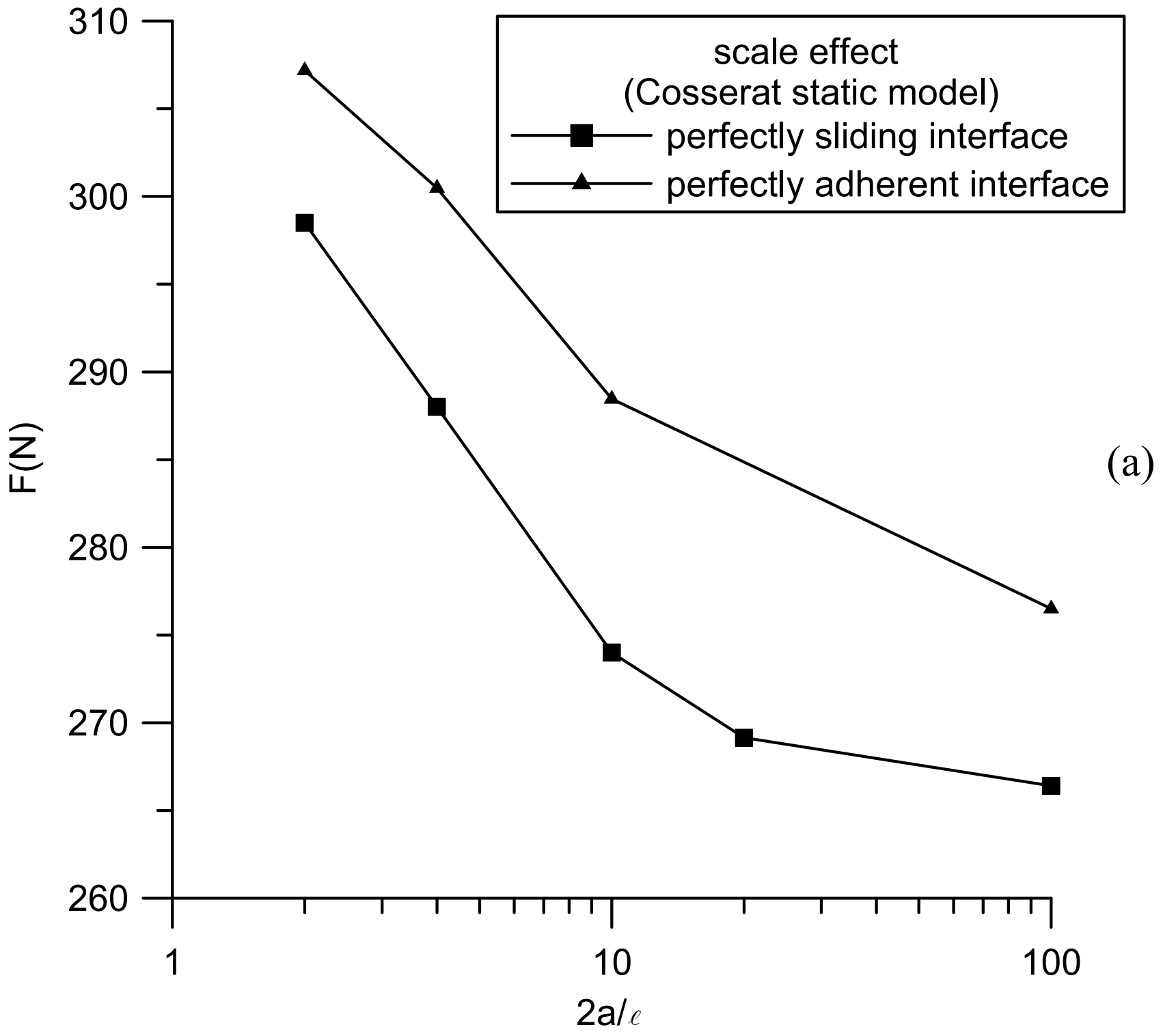

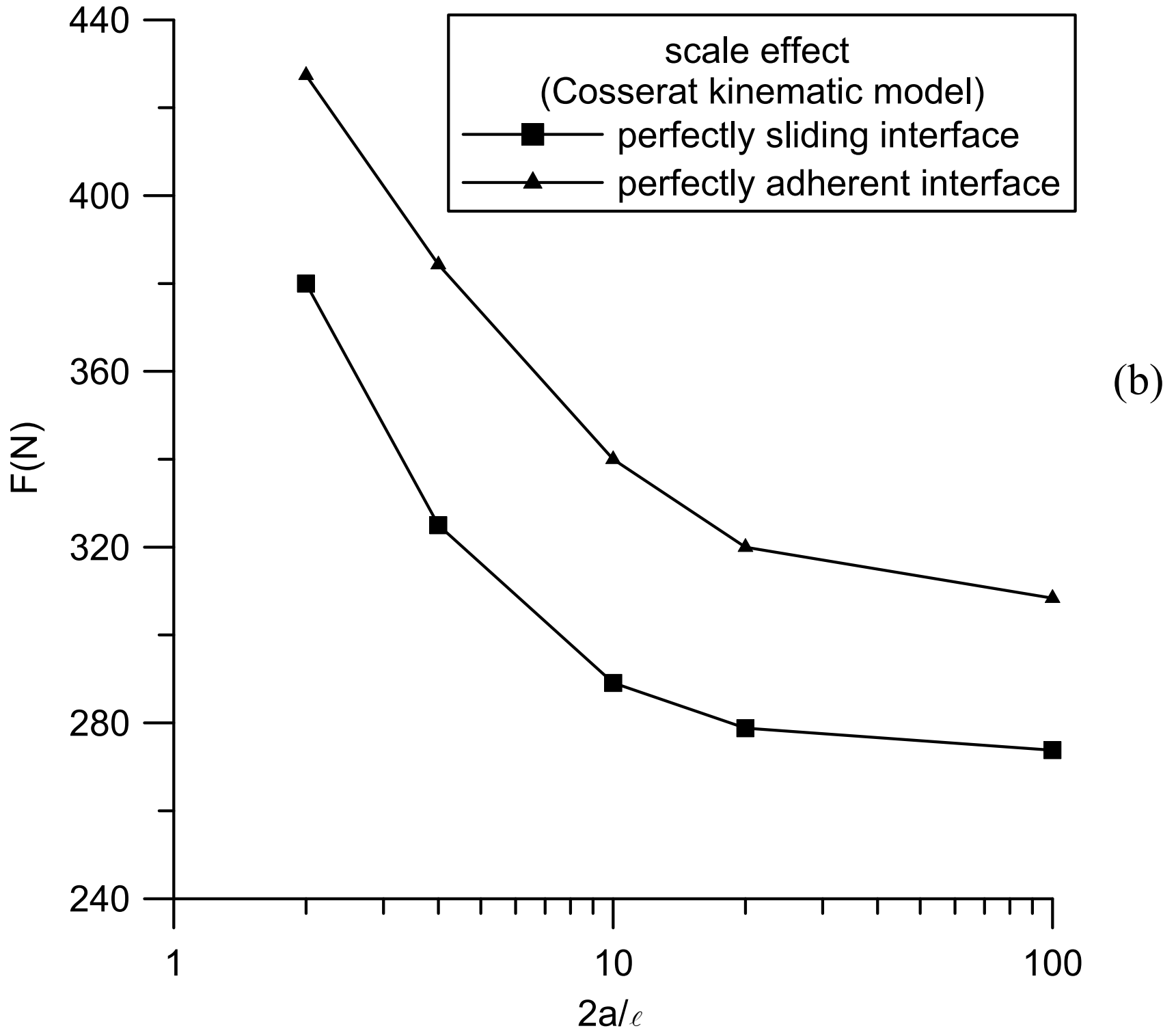

Figure 13: The effect of rock-tool interface friction on scale effect
(a) statical model, (b) kinematical model

| $\ell/2a$ | slope of the indentation curve ($N.mm^{-1}$) | |
|---|---|---|
| | Mesh M1 | Mesh M2 |
| 0 | 20150 | 19677 |
| 0.1 | 21018 | 20596 |
| 0.25 | 22150 | 21750 |
| 0.5 | 23665 | 23243 |

Table 1: Indentation of a Cosserat elastic half-space (frictionless tool): Slope of indentation curve